\documentclass{webofc}

\usepackage[varg]{txfonts}   
\usepackage{hyperref}
\usepackage{url}
\hypersetup{colorlinks=true,citecolor=blue,urlcolor=blue,linkcolor=blue}
\begin{document}
\title{Analysis of the semileptonic $B_c\to D_s^*\mu^+ \mu^-$ decay mode in the effective field theory approach}
%
%

\author{\firstname{Ajay Kumar} \lastname{Yadav}\inst{1}\fnsep\thanks{\email{yadavajaykumar286@gmail.com}}\and\firstname{Manas Kumar} \lastname{Mohapatra}\inst{2}\fnsep\thanks{\email{manasmohapatra12@gmail.com}} 
         \and
        \firstname{Suchismita} \lastname{Sahoo}\inst{1}\fnsep\thanks{\email{suchismita8792@gmail.com}}
}

\institute{Department of Physics, Central University of Karnataka, Kalaburagi-585367, India  \and School of Physics, University of Hyderabad, Hyderabad-500046, India}

\abstract{Discrepancies have been noted between experimental measurements and Standard Model predictions for various observables related to the $B\to(K, K^*,\phi) ll$ processes.  Recently, the Belle-II Collaboration observed a $2.8\sigma$ deviation from Standard Model predictions in the branching ratio of the  $B\to K\nu_l\Bar{\nu_l}$ decay mode.  In this context, we investigate the exclusive semileptonic $B_c\to D_s^{*}\mu^+ \mu^-$ decay mode, mediated by $b\to s$ quark level transition, using Effective Field Theory formalism.  We perform a global fit to the new parameters using existing experimental data on $b \to s \mu^+ \mu^-$.  We then estimate the branching ratio and other  observables such as forward-backward asymmetry, lepton polarisation asymmetry, and the lepton non-universality parameter of the $B_c\to D_s^{(*)}\mu^+ \mu^-$ process. Although this process has not yet been detected experimentally, we offer predictions and discuss the observables for  $B_c\to D_s^{*}\mu^+ \mu^-$ in both the Standard Model and potential new physics scenarios.
}
\maketitle
\section{Introduction}
\label{intro}
The Standard Model (SM) is a remarkable framework for understanding fundamental particles and their interactions. However, it does not provide a complete picture of the Universe, such as the matter-antimatter asymmetry, gravitational interaction, neutrino mass mechanisms, and the nature of dark matter and dark energy.  Due to the disagreements between the SM predictions and experimental observations, the $b\to sll$ transition of $B$ meson decays has been a hot topic of discussion. This transition involves flavor-changing neutral currents, which are not allowed at the tree level in the SM and these currents are further suppressed by elements of the Cabibbo–Kobayashi–Maskawa (CKM) matrix, making them highly sensitive to small effects from new physics. Recently, there have been advancements in the field of physics with the simultaneous measurements of $R_K$ and $R_{K^*}$ with $q^2$ values $[0.1-1.1]$ and as well as $[1.1-6]$, demonstrating excellent agreement with SM predictions at $0.2\sigma$\cite{Seuthe:2023xek,LHCb:2022vje}. However, there are still discrepancies in other observables, such as  the branching fraction $BR(B\to\phi ll)$\cite{Buras:2022qip}. Additionally, the CMS experiment recently observed deviations in the angular observable $P^{'}_5(B\to K^{*}\mu^+\mu^-)$\cite{CMS:2017rzx,LHCb:2020gog,LHCb:2020lmf,Buras:2022qip,Alguero:2023jeh}. This discrepancies motivate scientists to conduct a diverse range of experiments to explore new aspects of physics and test the accuracy of current theories \cite{Woithe:2017lzd,Saikumar:2024ahz}.

In this work, we study the rare semileptonic decay channel, $B_c\to D^{*}_s\mu^+\mu^-$ with the Effective Field Theory (EFT) formalism.   This channel has been studied using various approaches such as the relativistic quark model (RQM)\cite{Ebert:2010dv}, QCD sum rules\cite{Azizi:2008vv,Azizi:2008vy}, covariant confined quark model\cite{Ivanov:2019nqd,Ivanov:2024iat} the light front quark model, and the constituent quark model along with specific NP model like $Z'$ and Leptoquark\cite{Mohapatra:2021ynn} and in model independent approaches\cite{Dutta:2019wxo,Li:2023mrj}. Recently the LHCb Collaboration has established the upper limit for the $B_c\to D^{*}_s\mu^+\mu^-$ channel, as well as the fragmentation fractions of B meson with c and u quarks at a $95\%$ CL\cite{LHCb:2023lyb}.  We fit the new coefficients by using the existing data on $b \to s \mu^+ \mu^-$ for both the 1D and 2D scenarios. We then examine the sensitivities of each scenarios for the the branching ratio, forward backward asymmetry, lepton nonuniversality (LNU) parameter of $B_c\to D^{*}_s\mu^+\mu^-$ decay mode. 

The paper is structured in the following manner. In section 2, we delve into the theoretical framework of $B_c\to D^{*}_s\mu^+\mu^-$ decay process in the  EFT approach. We provide the global fit analysis for both the 1D and 2D scenarios of new Wilson coefficients in section 3.  The discussion in section 4 includes the effects of the constrained new parameters on the various observable of the $B_c\to D^{*}_s\mu^+\mu^-$ process. Section 5 encompasses the summary and conclusion.

\section{Theoretical Framework}

\subsection{General Effective Hamiltonian}
The most general effective Hamiltonian for the rare semileptonic decay $b\to sll$, can be written as\cite{Ahmed:2011sa},
\begin{align}
\centering
   \mathcal{H}_{\rm eff}&=-\frac{4\,G_F}{\sqrt{2}}V_{tb}\,V^*_{ts}\Big[C^{eff}_7\,\mathcal{O}_7+C_7^{'}\,\mathcal{O}_7^{'}+\sum_{i=9,10,P,S}((C_i+C_i^{NP})\,\mathcal{O}_i+C_i^{'NP}\,\mathcal{O}_i^{'})\Big]\,, \label{eff-ham}
\end{align}
where $G_F$ is the Fermi constant, $V_{ij}$ are the CKM matrix elements,  $\mathcal{O}_i^{(')}$ are the effective operators
\begin{eqnarray*}
     \mathcal{O}_7&&=\frac{e}{16\,\pi^2}\,m_b\,(\bar{s}\sigma_{\mu\,\nu}P_R\,b)\,F^{\mu\,\nu},~~~~~
     \mathcal{O}^{'}_7=\frac{e}{16\,\pi^2}\,m_b\,(\bar{s}\sigma_{\mu\,\nu}P_L\,b)\,F^{\mu\,\nu}\\
     \mathcal{O}_9&&=\frac{e^2}{16\,\pi^2}\,(\bar{s}\gamma_{\mu}P_L\,b)\,(\bar{\mu}\gamma^{\mu}\mu),~~~~~~~
     \mathcal{O}^{'}_9=\frac{e^2}{16\,\pi^2}\,(\bar{s}\gamma_{\mu}P_R\,b)\,(\bar{\mu}\gamma^{\mu}\mu),\\
      \mathcal{O}_{10}&&=\frac{e^2}{16\,\pi^2}\,(\bar{s}\gamma_{\mu}P_L\,b)\,(\bar{\mu}\gamma^{\mu}\,\gamma_5\,\mu),~~~~
     \mathcal{O}^{'}_{10}=\frac{e^2}{16\,\pi^2}\,(\bar{s}\gamma_{\mu}P_R\,b)\,(\bar{\mu}\gamma^{\mu}\,\gamma_5\,\mu),\\
      \mathcal{O}_S&&=\frac{e^2}{16\,\pi^2}\,m_b\,(\bar{s}P_R\,b)\,(\bar{\mu}\mu),~~~~~~~~~~
     \mathcal{O}^{'}_S=\frac{e^2}{16\,\pi^2}\,m_b\,(\bar{s}P_L\,b)\,(\bar{\mu}\mu),\\
     \mathcal{O}_P&&=\frac{e^2}{16\,\pi^2}\,m_b\,(\bar{s}P_R\,b)\,(\bar{\mu}\gamma_5\mu),~~~~~~~
     \mathcal{O}^{'}_P=\frac{e^2}{16\,\pi^2}\,m_b\,(\bar{s}P_L\,b)\,(\bar{\mu}\gamma_5\mu),\\
\end{eqnarray*}
and $C_i^{(')}$ are corresponding Wilson coefficients. In the Standard Model, the primed coefficients are zero, but they can be nonzero in the presence of new physics.  The coefficient $C_9^{eff}$ includes a short-distance perturbative component, defined as:
 \begin{equation}
     C_9^{eff}(q^2)=C_9+Y_{SD}(q^2)
 \end{equation}
 where, the sort-distance contribution part is given as\cite{Ali:1999mm}
     \begin{eqnarray*}
         Y_{SD}(q^2)&=&h\big(\frac{m_c}{m_B},q^2\big)(3C_1+C_2+3C_3+C_4+3C_5+C_6)-\frac{1}{2}h(1,q^2)(4C_3+4C_4+3C_5+C_6)\\
         &&-\frac{1}{2}h(0,q^2)(C_3+3C_4)+\frac{2}{9}(3C_3+C_4+3C_5+C_6).\\
    h(0,q^2)&=&\frac{8}{27}-\ln\frac{m_b}{\mu}-\frac{4}{9}\ln s+\frac{4}{9}i\pi,\\
     h(z,q^2)&=&-\ln\frac{m_b}{\mu}-\frac{8}{9}\ln z+\frac{8}{27}\times\frac{4}{9}\times\frac{4z^2}{s}\\
     &&-\frac{2}{9}\Big(2+\frac{4z^2}{s}\Big)\sqrt{|1-\frac{4z^2}{s}|}
     \begin{cases}
     \ln\Big|\frac{\sqrt{1-\frac{4z^2}{s}}+1}{\sqrt{1-\frac{4z^2}{s}}-1}\Big|-i\pi,&\text{for}\frac{4z^2}{s} < 1 \\
     2 \arctan \frac{1}{\sqrt{\frac{4z^2}{s}-1}},&\text{for}\frac{4z^2}{s} > 1 
     \end{cases}
     \end{eqnarray*}
     with, $s=\frac{q^2}{m_B^2}$ and $ m_B$ is the mass of  $B$ meson.

     \subsection{$B_c\to D^{*}_s\mu^+\mu^-$}
     Using the general effective Hamiltonian \ref{eff-ham},  the $q^2$ dependent differential branching ratio for the $B_c\to D_s^*\mu^+\mu^-$ process is defined as 
     \begin{equation}
    \frac{d\Gamma}{dq^2}=\frac{1}{4}\Big[3\,I_1^c+6\,I_1^s-I_2^c-2\,I_2^s\Big],
\end{equation}
where $I_{1,2}^{c,s}$'s are the angular coefficients.  Besides the differential decay distribution, we also scrutinize other physical observables such as forward-backward asymmetry $A_{FB}$, longitudinal (transverse) polarization fraction $F_{L(T)}$, form factor independent angular observables $P_4^{'}$, $P_5^{'}$ and the lepton nonuniversality  ratio. These observables are defined as\cite{Altmannshofer:2008dz,Matias:2012xw,Sahoo:2015qha} follows:
\begin{itemize}
    \item \textbf{Forward-backward asymmetry}:
\begin{equation}
    A_{FB}(q^2)=\frac{3\,I_6}{3\,I_1^c+6\,I_1^s-I_2^c-2\,I_2^s}\,.
\end{equation}

    \item \textbf{Longitudinal and transverse polarization fractions}:
    \begin{equation}
    F_L(q^2)=\frac{3\,I_1^c-I_2^c}{3\,I_1^c+6\,I_1^s-I_2^c-2\,I_2^s}\,,~~~~~ F_T(q^2)=1-F_L(q^2)\,.
\end{equation}
\item \textbf{Lepton non-universality ratio}:
      \begin{equation}
    R_{D_s^*}(q^2)=\frac{d\Gamma(B_c\rightarrow D_s^*\mu^+\mu^-)/dq^2}{d\Gamma(B_c\rightarrow D_s^*\,e^+\,e^-)/dq^2}\,.
\end{equation}
\item \textbf{Form factor independent observable} $(P_{4,5}^{'})$:
\begin{equation}
  P_4^{'}=\frac{I_4}{\sqrt{-I_2^c\,I_2^s}},~~~~~~~  P_5^{'}=\frac{I_5}{2\sqrt{-I_2^c\,I_2^s}}\,.
\end{equation}
 \end{itemize}
 All the angular coefficients $I_{i}^{c,s}$'s ($i=1,2...$) in terms of the transversity amplitudes and the transversity amplitudes as a functions of the form factors and Wilson coefficients are given in the appendix A.

\section{Numerical Fits to the Model Parameters}
This paper scrutinize the impact of (axial)vector coefficients on $B_c \to D_s^* \mu^+ \mu^-$ decay mode. To determine the best fit values of the (axial)vector coefficients $C_{9, 10}^{(')NP}$, we perform a $\chi^2$ fit of  these coefficients to the existing data on $b \to s \mu^+ \mu^-$ observables using the flavio package \cite{Straub:2018kue}. We have used the following $b \to s \mu^+ \mu^-$ observables for our analysis. 
\begin{itemize}
    \item \textbf{Branching ratios}: Branching ratios of rare (semi)leptonic decay modes mediated by the $b \to s \mu^+ \mu^-$ transitions such as $B_s \to \mu^+ \mu^-$, $B^{+(0)} \to K^{+(0)} \mu^+ \mu^-$, $B^{+(0)} \to K^{* +(0)} \mu^+ \mu^-$ and $B_s \to \phi \mu^+ \mu^-$ in different $q^2$ bins.
    \item \textbf{Physical observables of} $\boldsymbol{B_{(s)} \to K^*(\phi) \mu^+ \mu^-}$: Forward backward asymmetries, longitudinal polarization asymmetries, form factor independent observables, CP averaged angular observbales, CP asymmetries in different $q^2$ bins. 
\end{itemize}
We have taken two possible hypothesis: (a) only one Wilson coefficients at a time (1D) and (b) two Wilson coefficients at a time (2D). Fitting the above observables, the bestfit values, $1\sigma$ range, pull$=\sqrt{\chi^2_{\rm SM}-\chi^2_{\rm best-fit}}$ values and the p-value ($\%$) for the 1D scenarios are given in Table \ref{tab-1} and for the 2D scenarios are provided in Table \ref{tab-2}. 
\begin{table}[htb]
\centering
\caption{Best-fit values $[1\sigma]$, pull and p values of different new physics scenarios of  single coefficient.}
\label{tab-1}       
\begin{tabular}{|l|llll|}
\hline
Scenario & Coefficient & Best fit value [$1\sigma$]  & Pull & p-value ($\%$)  \\\hline
S - I & $C_9^{\rm NP}$ & $ -1.227$ $[\substack{-0.959 \\ -1.363}]$    & 4.665 & 46.0 \\
S - II & $C_{10}^{\rm NP}$ & $ 0.456 $ $[\substack{0.555 \\ 0.252}]$& 2.823 & 22.0 \\
S - III & $C_{9}^{\rm 'NP}$ & $ 0.082 $ $[\substack{0.353 \\- 0.252}]$&   0.261 & 14.0\\ 
S - IV & $C_{10}^{\rm 'NP}$ & $ -0.134 $ $[\substack{-0.050 \\ -0.252}]$   &  1.085 & 13.0\\
S - V & $C_9^{\rm NP}=C_{10}^{\rm NP}$ &  $ 0.023 $ $[\substack{0.250 \\ -0.050}]$  & 0.158 &15.0 \\
S - VI & $C_9^{\rm NP}=-C_{10}^{\rm NP}$ &   $ -0.971 $  $[\substack{-0.757 \\ -1.161}]$&   4.921 & 53.0 \\
S - VII & $C_{9}^{\rm 'NP}=C_{10}^{\rm 'NP}$ &   $ -0.135 $ $[\substack{-0.052 \\ -0.251}]$&   1.003 & 14.0  \\
S - VIII & $C_{9}^{\rm 'NP}=-C_{10}^{\rm 'NP}$ &  $ 0.109 $ $[\substack{0.147 \\ 0.048}]$&   1.046 & 15.0 \\
S - IX & $C_9^{\rm NP}=-C_{9}^{\rm 'NP}$ & $ -0.835 $ $[\substack{-0.656 \\ -0.959}]$&   3.993 & 31.0 \\
S - X & $C_9^{\rm NP}=-C_{10}^{\rm NP}=-C_{9}^{\rm 'NP}=-C_{10}^{\rm 'NP}$   & $ -0.374 $ $[\substack{-0.254 \\ -0.451}]$     & 3.067 & 24.0 \\
S - XI & $C_9^{\rm NP}=-C_{10}^{\rm NP}=C_{9}^{\rm 'NP}=-C_{10}^{\rm 'NP}$  & $-0.281$ $[\substack{-0.150 \\ -0.454}]$&   2.415  & 20.0 \\\hline
\end{tabular}
\end{table}
\begin{table}[htb]
\centering
\caption{Best-fit values $[1\sigma]$, pull and p values of different new physics scenarios of two coefficients.}
\label{tab-2}       
\begin{tabular}{|l|llll|}
\hline
Scenario & Coefficient & Best fit value [$1\sigma$]  & Pull & p-value ($\%$)  \\\hline
S - I & $(C_9^{\rm NP},C_{10}^{\rm NP})$ & $(-1.398 [\substack{-1.219 \\ -1.578}], 0.694[\substack{0.855 \\ 0.480}])$ &  5.961& 67.0 \\ 
S - II &  $(C_9^{\rm NP},C_{9}^{\rm 'NP})$ &  $(-1.206 [\substack{-0.988 \\ -1.428}]$, $-0.053 [\substack{0.301 \\ -0.386}])$ & 4.640 & 47.0 \\ 
S - III &  $(C_9^{\rm NP},C_{10}^{\rm 'NP})$ &  ($-1.269[\substack{-1.079 \\ -1.454}]$, $-0.306[\substack{-0.158 \\ -0.390}]$ )  & 5.239 & 53.0\\ 
S - IV & $(C_{10}^{\rm NP},C_{9}^{\rm 'NP})$ & ($0.516[\substack{0.634 \\ 0.291}] $, $ -0.050 [\substack{0.333 \\ -0.341}]$) & 3.106 & 16.0\\ 
S - V & $(C_{10}^{\rm NP},C_{10}^{\rm 'NP})$ & ($0.469[\substack{0.680 \\ 0.262}] $, $ 0.101 [\substack{0.237 \\ -0.088}]$)   & 2.565 & 20.0 \\ 
S - VI & $(C_{9}^{\rm 'NP},C_{10}^{\rm 'NP})$ &  ($-0.033[\substack{0.283 \\ -0.379}] $, $ -0.133 [\substack{-0.008 \\ -0.278}]$)  & 1.059 & 13.0 \\ 
S - VII & $(C_{9}^{\rm NP}=-C_{9}^{\rm 'NP},C_{10}^{\rm NP}=C_{10}^{\rm 'NP})$ &  ($-0.811[\substack{0.636 \\ -0.996}] $, $ 0.128 [\substack{0.273 \\ -0.018}]$) & 3.689 & 36.0\\ 
S - VIII & $(C_{9}^{\rm NP}=C_{9}^{\rm 'NP}, C_{10}^{\rm NP}=-C_{10}^{\rm 'NP})$ &  ($-1.121[\substack{-0.911 \\ -1.330}] $, $ 0.302 [\substack{0.365 \\ 0.198}]$)  & 5.174 & 51.0 \\ 
S - IX & $(C_{9}^{\rm NP}=C_{9}^{\rm 'NP},C_{10}^{\rm NP}=C_{10}^{\rm 'NP})$ &  ($-0.838[\substack{-1.077 \\ -0.600}] $, $ 0.015 [\substack{-0.141 \\ 0.172}]$) & 4.032 & 31.0\\ 
S - X & $(C_{9}^{\rm NP}=-C_{10}^{\rm NP},C_{9}^{\rm 'NP}=C_{10}^{\rm 'NP})$ & ($-0.986[\substack{-0.783 \\ -1.189}] $, $ 0.108 [\substack{0.232 \\ -0.015}]$)  & 5.412 & 53.0\\ 
S - XI & $(C_{9}^{\rm NP}=-C_{10}^{\rm NP},C_{9}^{\rm 'NP}=-C_{10}^{\rm 'NP})$ &  ($-1.008[\substack{-0.800 \\ -1.217}] $, $ -0.089 [\substack{0.033 \\ -0.211}]$)  & 4.922 & 49.0\\\hline
\end{tabular}
\end{table}
Table \ref{tab-1} shows that the pull value for the S-VI ($C_9^{\rm NP}=-C_{10}^{\rm NP}$) scenario is largest among all possible 1D scenarios, followed by the S - I ($C_9^{\rm NP}$) scenario. This indicates that these two scenarios provide the best and most acceptable fits compared to the others. Similarly, the $(C_9^{\rm NP},C_{10}^{\rm NP})$ combination of two Wilson coefficient has the maximum pull value $=5.961$, followed by S - X $(C_{9}^{\rm NP}=-C_{10}^{\rm NP},C_{9}^{\rm 'NP}=C_{10}^{\rm 'NP})$ scenario, S - III  $(C_9^{\rm NP},C_{10}^{\rm 'NP})$ and S - VIII  $(C_{9}^{\rm NP}=C_{9}^{\rm 'NP}, C_{10}^{\rm NP}=-C_{10}^{\rm 'NP})$ scenario. 

\section{Results and Discussion}
\label{lab-4}

\begin{figure}[htb]
\begin{center}
\includegraphics[height=37mm,width=60mm]{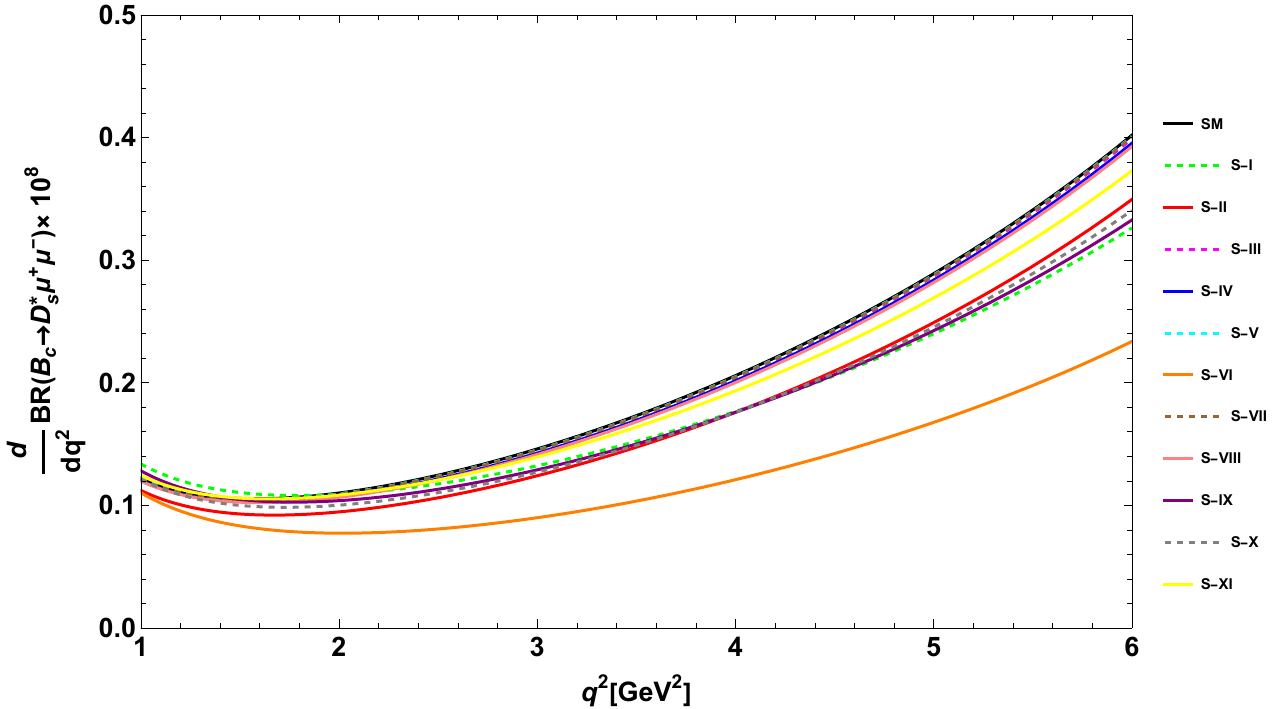}
\includegraphics[height=37mm,width=60mm]{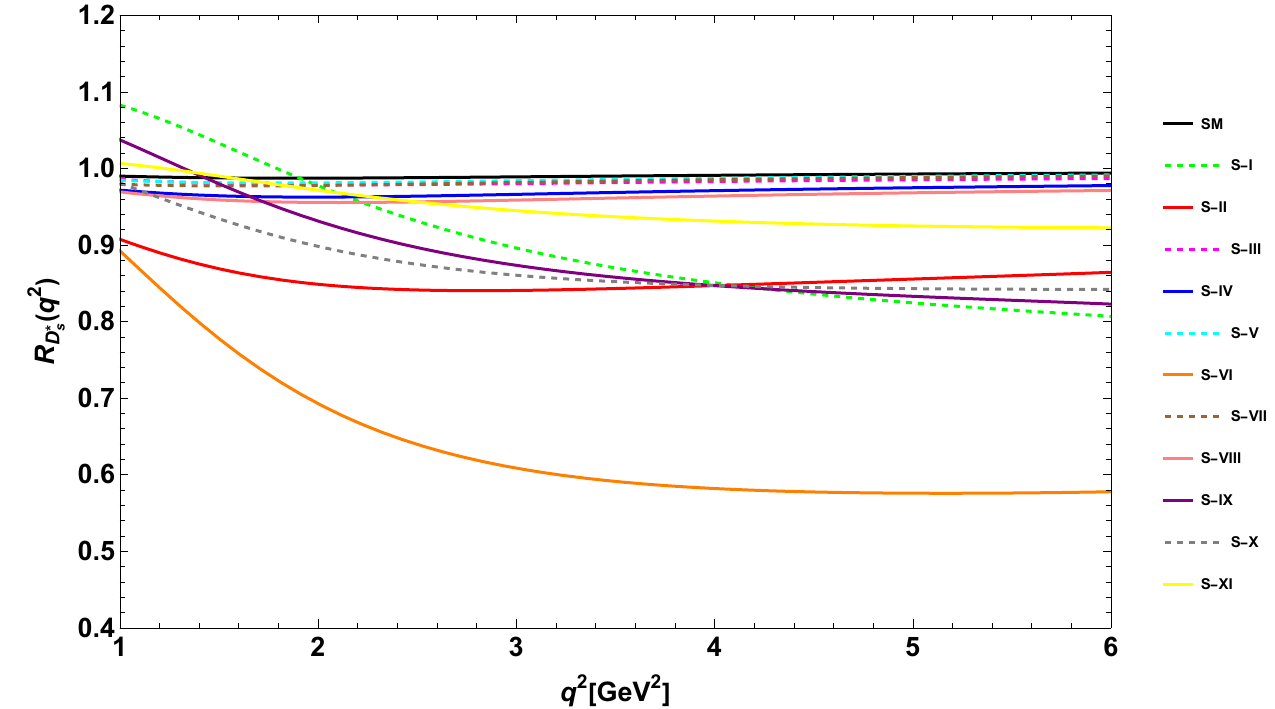}
\includegraphics[height=37mm,width=60mm]{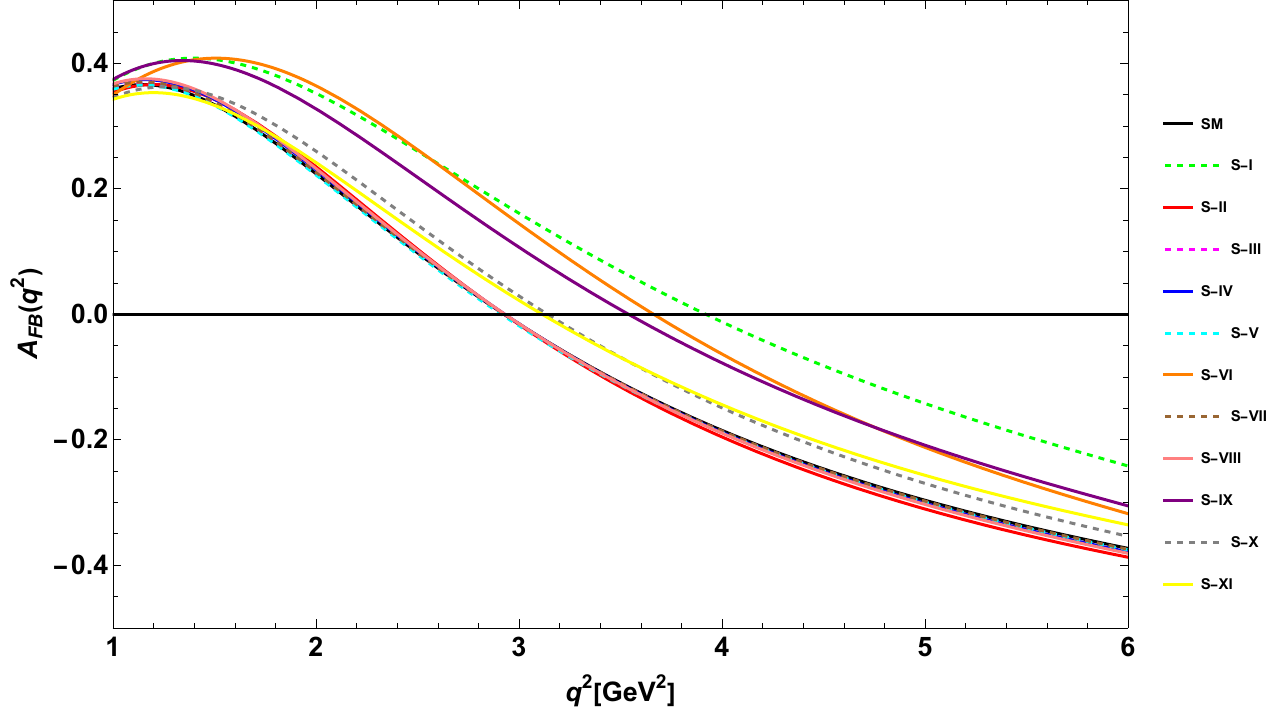}
\includegraphics[height=37mm,width=60mm]{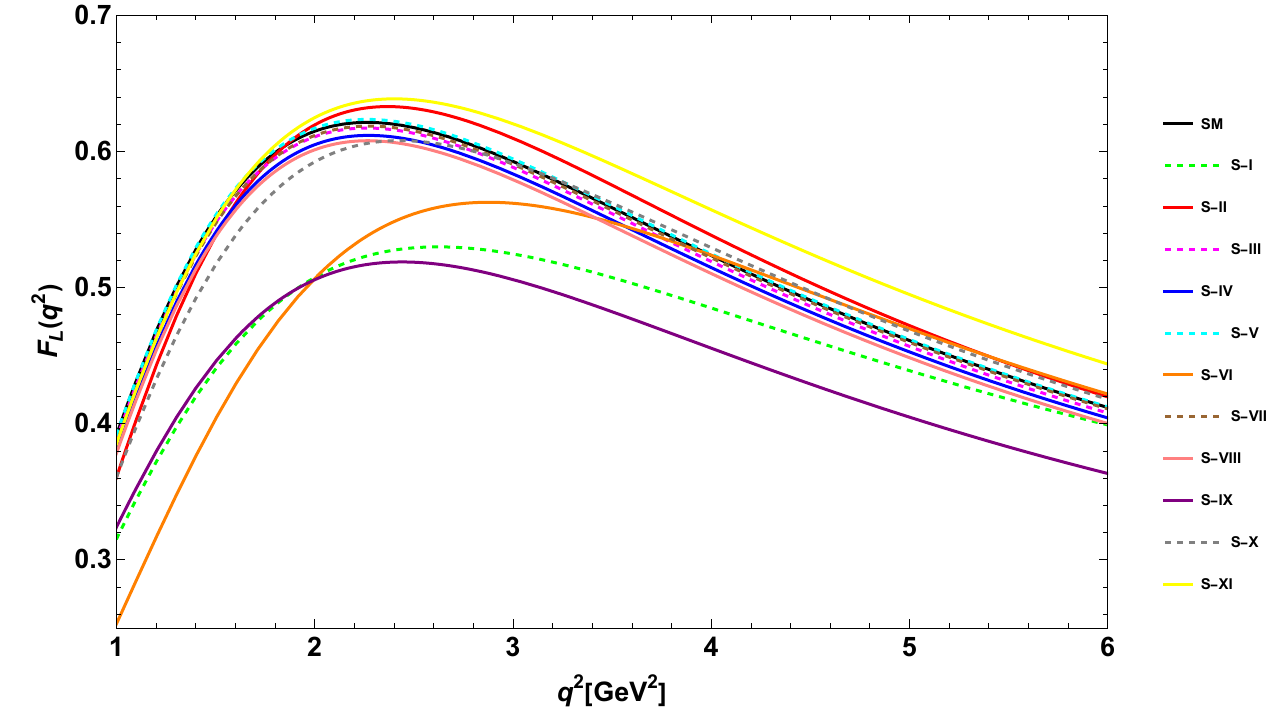}
\includegraphics[height=37mm,width=60mm]{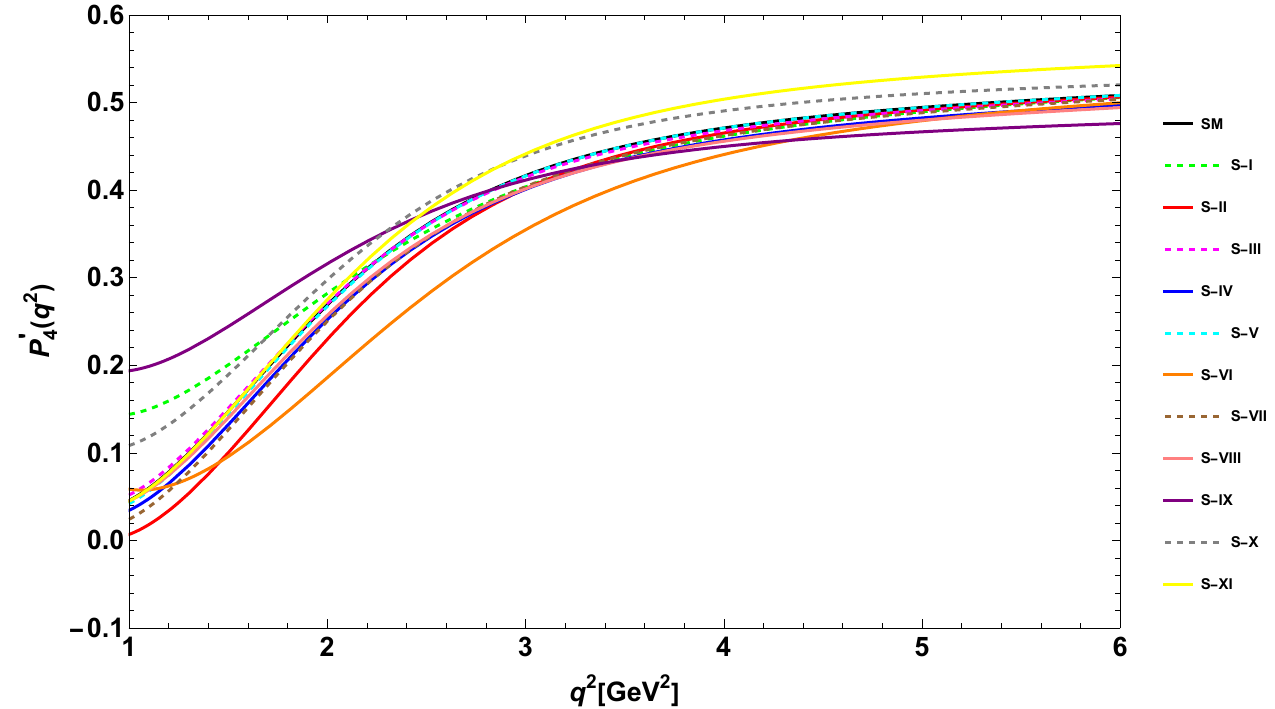}
\includegraphics[height=37mm,width=60mm]{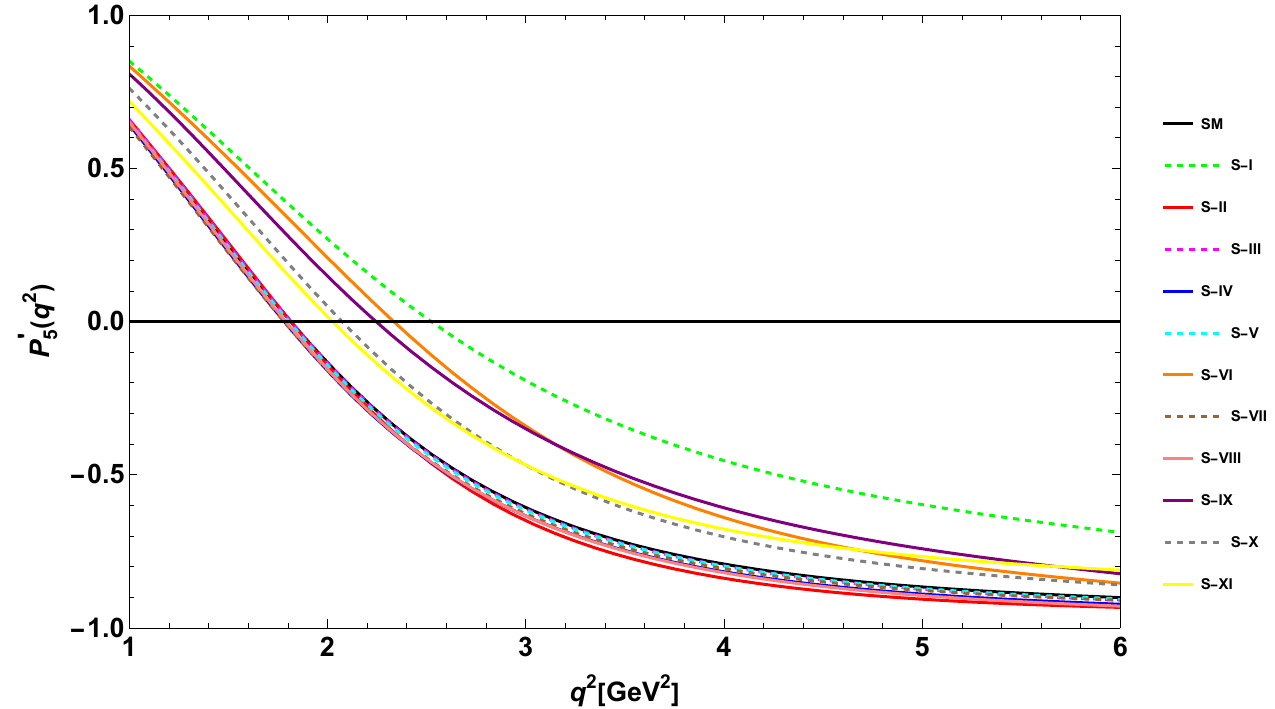}
\caption{The $q^2 \in [1-6]$ behavior of branching ratio (top left),  $R_{D_s^*}$ (top right), $A_{FB}$ (middle left), $F_L$ (middle right),  $P_4^{'}$ (bottom left) and $P_5^{'}$ (bottom right) observables of $B_c \to D_s^* \mu^+ \mu^-$ with 1D scenarios. Here, the black solid lines represent the Standard Model predictions, while the solid and dashed lines of different colors represent results for various scenarios. } \label{fig:1D}
\end{center}
\end{figure}

\begin{figure}[htb]
\begin{center}
\includegraphics[height=37mm,width=60mm]{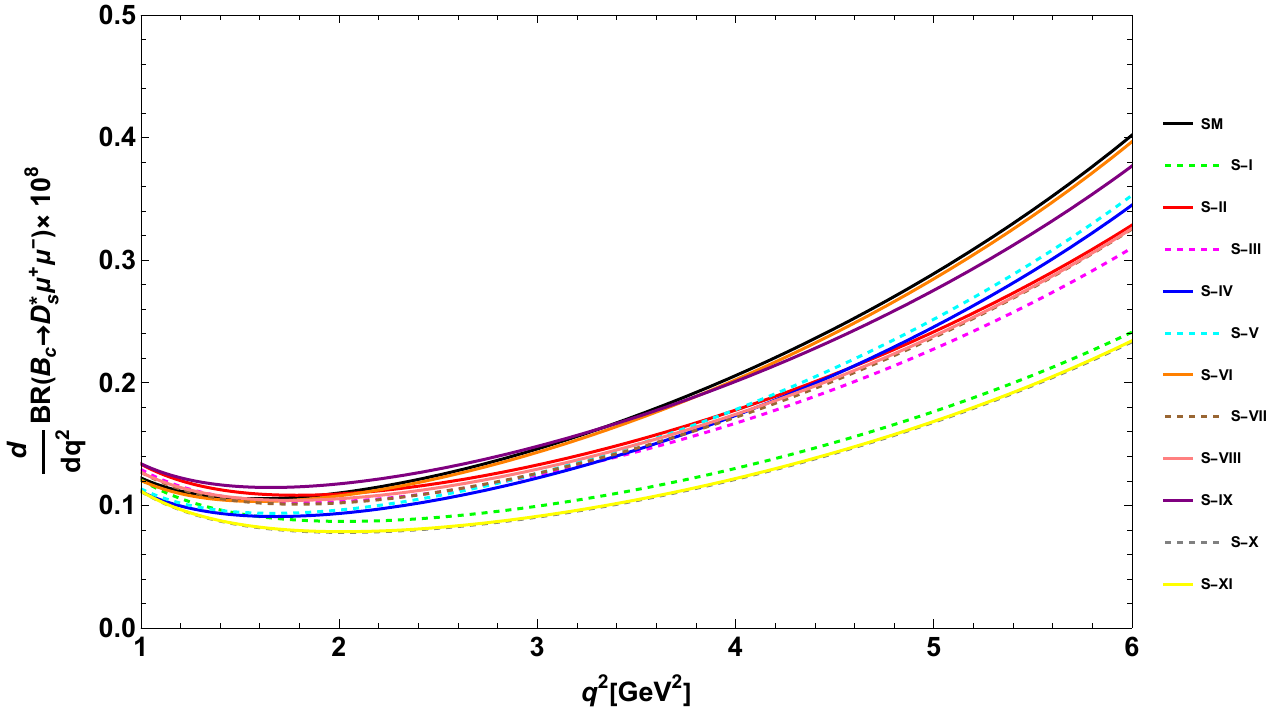}
\includegraphics[height=37mm,width=60mm]{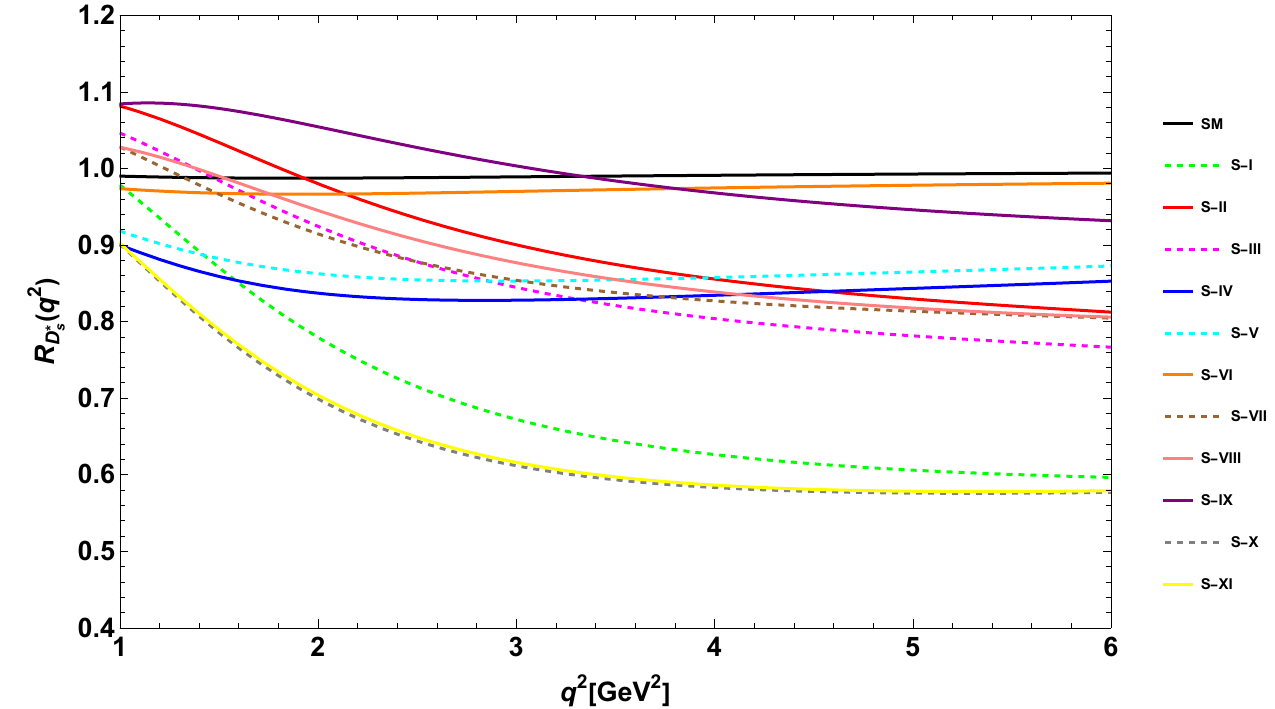}
\includegraphics[height=37mm,width=60mm]{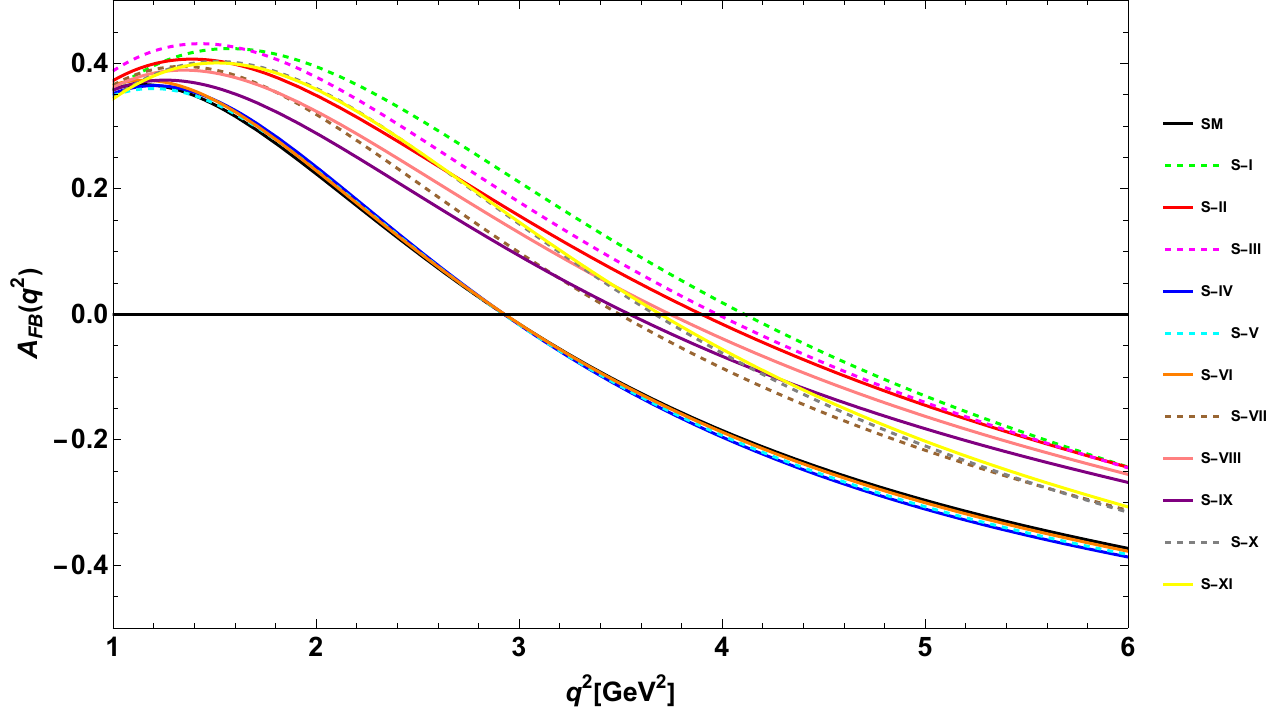}
\includegraphics[height=37mm,width=60mm]{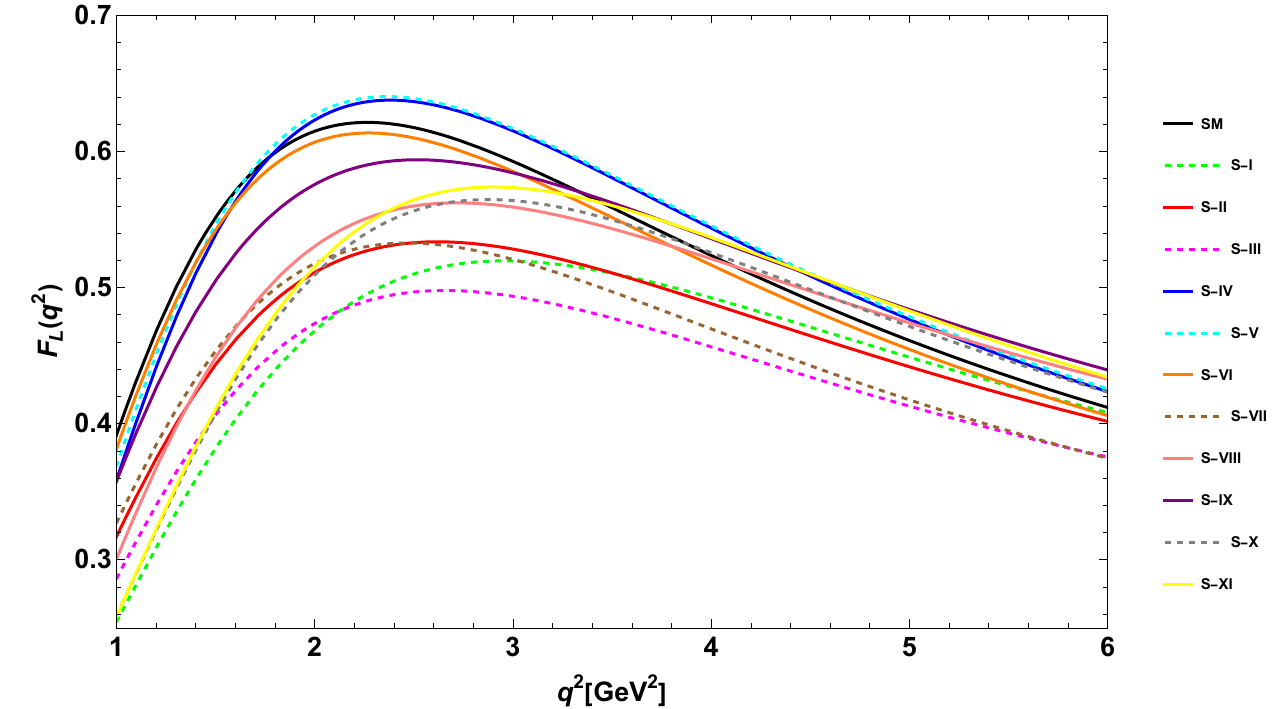}
\includegraphics[height=37mm,width=60mm]{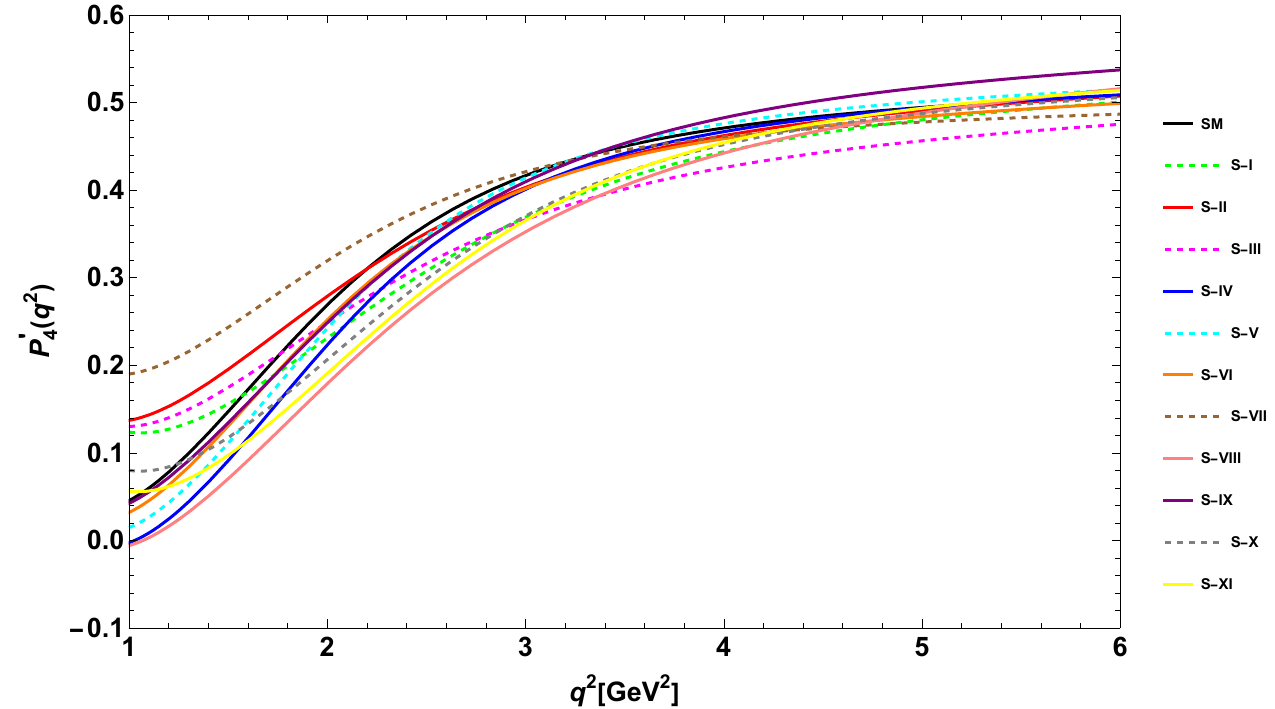}
\includegraphics[height=37mm,width=60mm]{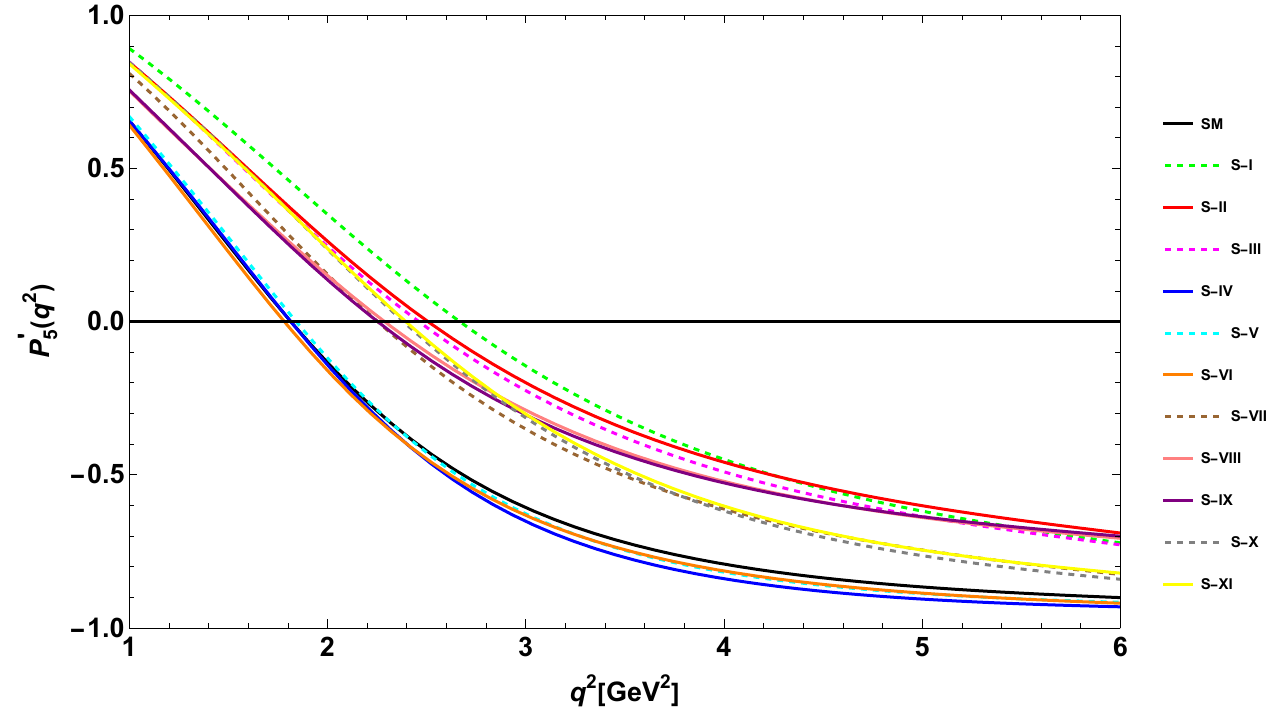}
\caption{The $q^2$ behavior of branching ratio (top left),  $R_{D_s^*}$ (top right), $A_{FB}$ (middle left), $F_L$ (middle right),  $P_4^{'}$ (bottom left) and $P_5^{'}$ (bottom right) observables of $B_c \to D_s^* \mu^+ \mu^-$ with 2D scenarios. The black solid lines represent Standard Model predictions, while the colored solid and dashed lines denote results for various scenarios.} \label{fig:2D}
\end{center}
\end{figure}

After collecting the best-fit values of all possible 1D and 2D scenarios of new coefficients, we now proceed to examine the sensitivity of these coefficients on the branching ratio and various angular observables of $B_c \to D_s^* \mu^+ \mu^-$ decay process. Using the best-fit values of all the 1D scenarios, the $q^2 \in [1,6]$ variation of the branching ratio (top left), $R_{D_s^*}$ (top right), forward-backward asymmetry (middle left), longitudinal polarization asymmetry (middle right), $P_4^{\prime}$ (bottom left) and $P_5^{\prime}$ (bottom right) of $B_c \to D_s^* \mu^+ \mu^-$ decay mode is presented in the Fig. \ref{fig:1D}.
In these figures, the black solid lines denote predictions from the Standard Model, while the colored solid and dashed lines represent results for various scenarios involving one coefficient. The branching ratio and other observables show the maximum deviation from the SM for the scenarios S-VI ($C_9^{\rm NP}=-C_{10}^{\rm NP}$) and S-I ($C_9^{\rm NP}$). When only the new coefficient $C_9^{\rm NP}$ is present, the zero crossings of $A_{FB}$ and $P_5^{'}$ observables exhibit the greatest deviation. The  predicted numerical values for $q^2 \in [1,6]$ bin are provided in Table \ref{tab-3}. 
\begin{table}[htb]
\centering
\caption{Numerical predictions for  observables of $B_c\to D_s^*\mu^+\mu^-$   with 1D scenarios.}
\label{tab-3}       
\begin{tabular}{|l|llllll|}
\hline
Scenarios & $BR\times10^{8}$ &$<R_{D_s^*}>$& $<A_{FB}>$ &$<F_L>$ & $<P_4^{'}>$& $<P_5^{'}>$ \\\hline
SM&$0.995909$&$0.990933$&$-0.15029$&$0.502247$&~0.43307~&$-0.665084$  \\
S - I&$0.872324$&$0.867965$&$0.0141311$&$0.458742$&0.425605&$-0.324654$ \\
S - II&$0.858233$&$0.853944$&$-0.157366$&$0.51103$&0.421312&$-0.697741$ \\
S - III&$0.988079$&$0.983141$&$-0.151481$&$0.497999$&0.430796&$-0.67033$ \\
S - IV&$0.976067$&$0.97119$&$-0.152712$&$0.493239$&0.419947&$-0.689361$ \\
S - V&$0.991937$&$0.98698$&$-0.153195$&$0.503311$&0.43277&$-0.671721$ \\
S - VI&$0.613327$&$0.610262$&$-0.0277296$&$0.47869$&0.390046&$-0.46957$ \\
S - VII&$0.989827$&$0.984881$&$-0.151105$&$0.500223$&0.423805&$-0.681056$ \\
S - VIII&$0.969369$&$0.964525$&$-0.15354$&$0.489112$&0.41931&$-0.691372$ \\
S - IX&$0.865568$&$0.861242$&$-0.045553$&$0.433772$&0.421539&$-0.469868$ \\
S - X&$0.863225$&$0.858911$&$-0.113459$&$0.502416$&0.449859&$-0.559257$ \\
S - XI&$0.942656$&$0.937945$&$-0.11124$&$0.530621$&0.458189&$-0.545882$\\
\hline
\end{tabular}
\end{table}
Fig. \ref{fig:2D} illustrates the impact of 2D scenarios on various observables for the decay $B_c \to D_s^* \mu^+ \mu^-$ with $q^2$. The panels show the branching ratio (top left), $R_{D_s^*}$ (top right), forward-backward asymmetry (middle left), longitudinal polarization asymmetry (middle right), $P_4^{\prime}$ (bottom left) and $P_5^{\prime}$ (bottom right). The solid black lines represent SM results, while the colored solid and dashed lines depict predictions from various 2D scenarios. The deviation in almost all the observables are significant for S-I $(C_9^{\rm NP},C_{10}^{\rm NP})$, S-X $(C_{9}^{\rm NP}=-C_{10}^{\rm NP},C_{9}^{\rm 'NP}=C_{10}^{\rm 'NP})$ and S-XI $(C_{9}^{\rm NP}=-C_{10}^{\rm NP},C_{9}^{\rm 'NP}=-C_{10}^{\rm 'NP})$ scenarios. Scenarios S-I and S-III induce the largest shifts in the zero crossings of the $A_{FB}$ and $P_5^{'}$ observables. The predited numerical values for 2D scenarios are presented in Table \ref{tab-4}.
\begin{table}[htb]
\centering
\caption{Numerical predictions for  observables of $B_c\to D_s^*\mu^+\mu^-$  with 2D scenarios.}
\label{tab-4}       
\begin{tabular}{|l|llllll|}
\hline
Scenario & $BR\times10^{8}$ &$<R_{D_s^*}>$& $<A_{FB}>$ &$<F_L>$ & $<P_4^{'}>$ & $<P_5^{'}>$\\\hline
SM&$0.995909$&$0.990933$&$-0.15029$&$0.502247$&0.43307&$-0.665084$  \\
S - I&$0.659425$&$0.65613$&$0.0450635$&$0.452897$&0.400632&$-0.298963$ \\
S - II&$0.876757$&$0.872375$&$0.0108972$&$0.461804$&0.425682&$-0.33026$ \\
S - III&$0.826409$&$0.822279$&$0.0233328$&$0.430211$&0.392443&$-0.358219$ \\
S - IV&$0.846171$&$0.841942$&$-0.157336$&$0.515268$&0.421405&$-0.697981$ \\
S - V&$0.868733$&$0.864392$&$-0.15552$&$0.517975$&0.431423&$-0.678396$ \\
S - VI&$0.979493$&$0.974599$&$-0.152309$&$0.495055$&0.421005&$-0.687147$ \\
S - VII&$0.846813$&$0.842581$&$-0.0530879$&$0.446196$&0.430432&$-0.470705$ \\
S - VIII&$0.858104$&$0.853815$&$-0.0113529$&$0.489383$&0.395102&$-0.397493$ \\
S - IX&$0.981172$&$0.976268$&$-0.0393625$&$0.509791$&0.43836&$-0.406841$ \\
S - X&$0.615151$&$0.612077$&$-0.026341$&$0.480789$&0.401619&$-0.447669$ \\
S - XI&$0.618286$&$0.615196$&$-0.0213378$&$0.490384$&0.400763&$-0.434288$\\
\hline
\end{tabular}
\end{table}

\section{Conclusion}
In this work, we studied the $B_c\to D_s^*\mu^+\mu^-$ decay mode using an Effective Field Theory approach, considering cases with either one (1D) or two (2D) unknown Wilson coefficients associated with (axial)vector operators. We performed a global fit analysis of these new Wilson coefficients using existing experimental data on  $b\to s\mu^+\mu^-$ observables. Based on the best-fit values from both 1D and 2D scenarios, we estimated the branching ratio and various angular observables for $B_c\to D_s^*\mu^+\mu^-$ process in the $q^2$ range $ [1,6]~{\rm GeV}^2$. These observables include forward-backward asymmetry, longitudinal polarization asymmetry, form factor-independent observables, and lepton nonuniversality parameter. We observed significant deviations from the Standard Model predictions, particularly for the scenarios with only $C_9^{\rm NP}$ and  $C_9^{\rm NP} = -C_{10}^{\rm NP}$ in the 1D case, and for the 2D scenarios:  $(C_9^{\rm NP},C_{10}^{\rm NP})$,  $(C_{9}^{\rm NP}=-C_{10}^{\rm NP},C_{9}^{\rm 'NP}=C_{10}^{\rm 'NP})$ and  $(C_{9}^{\rm NP}=-C_{10}^{\rm NP},C_{9}^{\rm 'NP}=-C_{10}^{\rm 'NP})$. This analysis suggests the need for a thorough experimental investigation of the rare semileptonic decay $B_c\to D_s^*\mu^+\mu^-$ in the near future. 

\section*{Acknowledgment}
AKY expresses gratitude to the DST-Inspire Fellowship division, Government of India, for financial support under ID No. IF210687. MKM would like to acknowledge financial support from the IoE PDRF at the University of Hyderabad.

\appendix
\section{ The angular coefficients $I_i$}
    The angular coefficients $I_i$ as a functions of the transversity amplitudes are defined as \cite{Altmannshofer:2008dz,Bobeth_2008,Kruger:2005ep},
\begin{eqnarray*}
        I_1^c&=&\Big(|A_L^0|^2+|A_R^0|^2\Big)+8\,\frac{m_l^2}{q^2}\,Re\Big[A_L^0\,A_R^{0*}\Big]+4\frac{m_l^2}{q^2}|A_t|^2,\\
        I_2^c&=&-\beta_l^2\Big(|A_L^0|^2+|A_R^0|^2\Big),\\
        I_1^s&=&\frac{(2+\beta_l^2)}{4}\Big[|A_L^{\perp}|^2+|A_L^{||}|^2+|A_R^{\perp}|^2+|A_R^{||}|^2\Big]+\frac{4\,m_l^2}{q^2}\,Re\Big[A_L^{\perp}A_R^{\perp\,*}+A_L^{||}A_R^{||\,*}\Big],\\
        I_2^s&=&\frac{1}{4}\beta_l^2\Big[|A_L^{\perp}|^2+|A_L^{||}|^2+|A_R^{\perp}|^2+|A_R^{||}|^2\Big],\\
        I_3&=&\frac{1}{2}\beta_l^2\Big[|A_L^{\perp}|^2-|A_L^{||}|^2+|A_R^{\perp}|^2-|A_R^{||}|^2\Big],\\
        I_4&=&\frac{1}{\sqrt{2}}\beta_l^2\Big[Re\Big(A_L^0\,A_L^{||\,*}\Big)+Re\Big(A_R^0\,A_R^{||\,*}\Big)\Big],\\
        I_5&=&\sqrt{2}\beta_l\Big[Re\Big(A_L^0\,A_L^{\perp\,*}\Big)-Re\Big(A_R^0\,A_R^{\perp\,*}\Big)\Big],\\
    \end{eqnarray*}
where the transversity amplitude, $A_{R,L}^{\perp,||,0}$ in terms of form factors are given by 
\begin{eqnarray*}
    A_{R,L}^{0}&=&-N_{D_s^{*}}\frac{1}{2\,m_{D_s^{*}}\sqrt{q^2}}\,\Bigg [(C_9^{eff}\pm C_{10})\Big[(m_{B_s}^2\,- m_{D_s^{*}}^2\,- q^2)(m_{B_s}\,+ m_{D_s^{*}})\,A_1 -\frac{\lambda(m_{B_c}^2,m_{D_s^*}^2,q^2)}{m_{B_s}\,+ m_{D_s^{*}}}\,A_2 \Big]\\
    && + 2 m_b C_7^{eff}\Big[(m_{B_s}^2,+ 3\,m_{D_s^*}^2,- q^2)\,T_2 - \frac{\lambda(m_{B_c}^2,m_{D_s^*}^2,q^2)}{m_{B_s}^2-m_{D_s^*}^2}\,T_3 \Big]\Bigg]\,,\\
    A_{R,L}^{\perp}&=&N_{D_s^*}\sqrt{2}\Big[(C_9^{eff}\pm C_{10})\frac{\sqrt{\lambda(m_{B_c}^2,m_{D_s^*}^2,q^2)}}{m_{B_s}+m_{D_s^*}}\,V+\frac{\sqrt{\lambda(m_{B_c}^2,m_{D_s^*}^2,q^2)}\,2m_b C_7^{eff}}{q^2}\,T_1\Big]\,,\\
    A_{R,L}^{||}&=&-N_{D_S^*}\sqrt{2}\Big[(C_9^{eff}\pm C_{10})(m_{B_s}+m_{D_s^*})A_1+\frac{2m_b\,C_7^{eff}(m_{B_s}^2-m_{D_s^*}^2)}{q^2}T_2\Big]\,,\\
    A_{R,L}^t&=&2\,N_{D_S^*}\,C_{10}\frac{\sqrt{\lambda(m_{B_c}^2,m_{D_s^*}^2,q^2)}}{\sqrt{q^2}}A_0,
\end{eqnarray*}
with 
\begin{equation*}
    N_{D_s^*}=\Bigg[\frac{G_F^2\,\alpha_e^2}{3.2^{10}\pi^5\,m_{B_s}^3}|V_{tb}\,V_{ts}^*|^2\,q^2\sqrt{\lambda(m_{B_c}^2,m_{D_s^*}^2,q^2)}\Big(1-\frac{4\,m_l^2}{q^2}\Big)^{1/2}\Bigg]^{1/2}\,.
\end{equation*}

\bibliography{reference}

\end{document}